\documentclass[hf]{ceurart}

\usepackage{fancyvrb}
\usepackage{fvextra}
\usepackage{caption}
\usepackage{subcaption}

\makeatletter
\let\blx@rerun@biber\relax
\makeatother

\usepackage[
  style=numeric,
  datamodel=software, 
  abbreviate=false,
  natbib=true,
  sorting=ydnt,
  backend=biber,
  bibencoding=utf8,
  giveninits=true,
  url=false,
  doi=true,
  maxcitenames=10,
  maxbibnames=100]{biblatex}
\usepackage{software-biblatex}
\begin{document}

\copyrightyear{2023}
\copyrightclause{Copyright for this paper by its authors.
  Use permitted under Creative Commons License Attribution 4.0 International (CC BY 4.0).}

\conference{Wikidata'23: Wikidata workshop at ISWC 2023}

\title{Bravo MaRDI: 
A Wikibase Powered Knowledge Graph on Mathematics}

\author[1]{Moritz Schubotz}[%
email=moritz.schubotz@fiz-karlsruhe.de,
url=https://schubotz.org,
]
\address[1]{FIZ Karlsruhe~-- Leibniz Institute for Information Infrastructure, Berlin, Germany}

\author[2]{Eloi Ferrer}[%
email=ferrer@zib.de,
]

\author[1]{Johannes Stegmüller}[%
email=johannes.stegmueller@fiz-karlsruhe.de,
]

\author[1]{Daniel Mietchen}[%
email=daniel.mietchen@fiz-karlsruhe.de,
]

\author[1]{Olaf Teschke}[%
email=olaf.teschke@fiz-karlsruhe.de,
]

\author[2]{Larissa Pusch}[%
email=pusch@zib.de,
]

\author[2]{Tim OF Conrad}[%
email=conrad@zib.de,
]

\DeclareFieldFormat{eprint:swmath}{%
\mkbibacro{SWMATH}\addcolon\addspace%
\ifhyperref
{\href{https://swmath.org/#1}{%
\(\langle\)swMATH\addcolon\nolinkurl{#1}\(\rangle\)%
\iffieldundef{eprintclass}
{}
{\addspace\texttt{\mkbibbrackets{\thefield{eprintclass}}}}}}
{\(\langle\)swMATH\addcolon\nolinkurl{#1}\(\rangle\)%
\iffieldundef{eprintclass}
{}
{\addspace\texttt{\mkbibbrackets{\thefield{eprintclass}}}}}
}

\address[2]{Zuse Institute Berlin, Berlin, Germany}

\begin{abstract}
Mathematical world knowledge is a fundamental component of Wikidata. However, to date, no expertly curated knowledge graph has focused specifically on contemporary mathematics. Addressing this gap, the Mathematical Research Data Initiative (MaRDI) has developed a comprehensive knowledge graph that links multimodal research data in mathematics. This encompasses traditional research data items like datasets, software, and publications and includes semantically advanced objects such as mathematical formulas and hypotheses.
This paper details the abilities of the MaRDI knowledge graph, which is based on Wikibase, leading up to its inaugural public release, codenamed Bravo, available on \url{https://portal.mardi4nfdi.de}.
\end{abstract}
\maketitle

This paper is submitted to the ISWC Wikidata Workshop 2023 novel resource track. 

\section{The One-Stop-Portal Vision} \label{sc.portal}

The Mathematical Research Data Initiative (MaRDI)~\cite{MaRDIProject} strives to bring the FAIR data principles \cite{WilkinsonDAA16} to life for mathematics and its research data (MathRD).
A critical aspect of this ambition involves creating a unified portal\footnote{Analogous initiatives name this type of service as 'Single Point of Access'~\cite{RenatSPA}.} that offers comprehensive access to all open research data within mathematics and related fields.

In this manuscript, we introduce the first Bravo release of the portal, now freely accessible through direct access or various APIs. Rather than constructing such a portal from scratch - a tactic adopted by parallel initiatives in Germany's National Research Data Initiative (NFDI) - we chose to utilize Wikimedia technology, predominantly MediaWiki, as a foundation, for several reasons:

\begin{itemize}
\item MediaWiki offers a thriving open-source environment with a myriad of free and open-source software.
\item A considerable user base exists for MediaWiki; many mathematicians are already acquainted with Wikipedia and, to some extent, Wikidata.
\item Upon the project's completion, we have the option to incorporate our data into Wikidata or another compatible platform.
\item Both the \href{https://meta.wikimedia.org/wiki/Wikimedia_Community_User_Group_Math}{Wikimedia mathematics community} and portal users, notably mathematicians based in Germany, mutually benefit from our advancements.
\item We seamlessly inherit knowledge from Wikipedia and Wikidata, painstakingly curated by hundreds of thousands of contributors over the past two decades.
\end{itemize}

However, this decision comes with challenges; becoming part of the Wikimedia movement necessitates embracing their work culture and guidelines and engaging with the community to achieve consensus on sustainable solutions.
Our contributions in this context are as follows:

\begin{enumerate}
\item We deliver a current ecosystem to house Wikibase and associated services on-premises.
\item We customize the general-purpose Wikibase infrastructure to cater to the specific needs of the mathematical community.
\item We have devised a mechanism for importing external data that aligns with the Wikidata data model.
\item We have actively enhanced the MediaWiki ecosystem by expanding and improving services such as QuickStatements, MediaWiki extensions like Math and MathSearch, and libraries like the Wikibase Integrator.
\item We have seeded the MaRDI KG with more than 100 million triples\footnote{Refer to the latest statistics from \url{https://portal.mardi4nfdi.de/wiki/Portal}.} aggregated from various sources.
\end{enumerate}

This paper is organized as follows: Section~\ref{sc.tech} begins with an overview of our technology stack. This is followed by Section~\ref{sc.math}, which presents our mathematics-specific extensions. Subsequently, Section~\ref{sc.seed} details the infrastructure for seeding and updating the knowledge graph. In Section~\ref{sc.examples}, we illustrate the functionality of our system with an example. Finally, Section~\ref{sc.outlook} provides a conclusion and discusses future prospects.

\section{Technical Infrastructure Overview} \label{sc.tech}

The MaRDI portal's infrastructure was initially constructed using the \href{https://github.com/wmde/wikibase-release-pipeline/tree/wmde.6}{Wikibase release pipeline}. The release pipeline incorporates a docker-compose file that delineates several services, including MediaWiki with Wikibase extensions, the Wikidata Query Service, and the Blazegraph Knowledge Graph backend. The docker-compose file allows for the launching of a pre-set group of services via a single command, thus optimizing reproducibility and system or web host independence.

The MaRDI portal operates on a solitary open-stack instance with 32 GB of main memory and 8 CPUs, alongside a 300 GB HDD. Only Docker and Git are installed on the host system. All configurations are housed in partially encrypted Git repositories, which feature a CI pipeline enabling direct deployment to production without necessitating access to the actual machine. If the data volume increases, we have the capability to transition from a single-node deployment to a Docker swarm with minimal modifications to our service specifications.

The following provides a concise description of selected services from our infrastructure, which currently comprises \href{https://github.com/MaRDI4NFDI/portal-compose/blob/main/docker-compose.yml}{28 services}~\cite{docker-compose}:

\begin{description}
\item[Traefik] As the primary entry point to our portal, all external network traffic is channeled through the Traefik reverse proxy. Traefik is responsible for managing SSL certificates properly via in-built lets-encrypt management and protective measures to guard individual services against unauthorized use. Additionally, Traefik collates access logs, enabling centralized usage examinations for our interconnected services.
\item[MediaWiki] In order to maintain compatibility with Wikipedia and manage only one version of mathematics-specific extensions, we chose to fork the infrastructure and update the container with the same versions utilized by Wikipedia in production starting from a previous effort~\cite{mwdockerbase}.
\item[WDQS] WDQS is the abbreviation of Wikidata Query Service which we have adapted to query the MaRDI portal and renamed it to MaRDI Query Service. This service is deployed based on four Docker containers: the Blazegraph database storing the knowledge graph, an updater that synchronizes the Wikibase entries with the Blazegraph backend, a frontend GUI to perform SPARQL queries, and a proxy to protect the SPARQL endpoint. The GUI for this service can be accessed at \url{query.portal.mardi4nfdi.de}.
\item[LaTeXML] LaTeXML is a service that converts LaTeX input to content and presentation MathML. It is employed for MathSearch.
\item[BaseX] BaseX is an XML database utilized for formula search. The MathSearch extension is responsible for keeping this service current.
\item[Backup] A straightforward, custom backup process was installed to preserve precisely the data we wish to retain, including the information saved on the SQL database, the MediaWiki pages as XML and the uploaded files.
\item[Scholia] Scholia is a tool that facilitates scholarly exploration of subsets of Wikidata's knowledge graph via a web browser~\cite{nielsenF2017scholia}. Its backend queries the Wikidata Query Service through predefined SPARQL 
queries
parametrized by Wikidata identifiers, and its frontend displays the results of several of these queries as profiles. A multitude of such profile types exist, including works, venues, topics, authors, awards, events, or organizations. MaRDI's instance\footnote{\url{https://scholia.portal.mardi4nfdi.de/}} is deployed via a fork~\cite{scholiaMaRDI} of the original~\cite{scholia} deployed on the Wikimedia Toolforge\footnote{\url{https://scholia.toolforge.org/}}. The intent behind the fork is to ensure maximum customizability for mathematics, with general improvements fed back upstream wherever possible. The ongoing customization includes efforts towards federated queries that utilize both the MaRDI knowledge graph and Wikidata.
\item[Portainer] Portainer is a management tool for Docker containers, images, volumes, and networks that operates via a web interface. It provides multiple ways of interacting with containers, including viewing their logs and accessing the container console. Portainer is compatible with standalone Docker installations, Docker Swarm, and Kubernetes.
\item[Watchtower] Watchtower is a tool that keeps Docker images up to date and guarantees the early installation of critical patches, provided Docker containers follow semantic versioning.
\item[Prometheus \& Grafana] Prometheus is a monitor and alerting toolkit to monitor the health of the portal and the infrastructure. We visualize the data gathered by Prometheus using the dashboarding tool Grafana. 
\end{description}

To guard against unintended side effects arising from changes to this multifaceted service infrastructure, we implemented a \href{https://portal.mardi4nfdi.de/wiki/Project:Testing_concept}{Testing concept} that is triggered by any modification. This helps increase our confidence in not deploying erroneous code to our production environment.

While the technology to analyze usage of the portal based on the logs is available, we have not enabled it yet. However, to estimate the size of the user group, we compare our service with the usage data of zbMATH Open. zbMATH Open had approx. 950k unique visors in 2022 via the web interface and 22k unique API users/bots.

Additionally, Identity and Access Management (NFDIIAM) is a crucial aspect. MaRDI utilizes OAuth as a protocol, which is also supported by the NFDI's wide basic service, NFDIIAM.

\section{Mathematics-specific Extensions} \label{sc.math}

Our portal leverages two math-specific MediaWiki extensions, namely, Extension-Math~\cite{Schubotz2014mathoid,SchubotzS16,SchubotzGMTG20,ScharpfSG21} and Extension-MathSearch~\cite{schubotz2013making,Schubotz17,ScharpfSG18}. The former extension, deployed across all Wikimedia projects, including Wikipedia, facilitates the rendering of mathematical expressions. The latter makes mathematical expressions searchable and links knowledge graph data to HTML5.

\subsection{Rendering of Mathematical Expressions}

The MaRDI portal, alongside Wikipedia and Wikidata, employs texvc markup to express mathematical formulas.
This LaTeX variant does not permit runtime syntax changes, favoring context-free formula processing with a regular grammar.
Mathematical expressions in wikitext are denoted by the wikitext tag \texttt{<math>} (to be distinguished from the HTML5 element \texttt{<math>}).
These \texttt{<math>} tags enclose the texvc expressions, which are then rendered. 
In contrast, Wikibase introduces a data type specifically for mathematical expressions.
Examples of such notations can be found in the DLMF defining formula in the MaRDI-KG~\cite{CohlSMS15}\footnote{\url{https://portal.mardi4nfdi.de/wiki/Item:Q1799}} and in math-tagged elements in Wikitext\footnote{\url{https://portal.mardi4nfdi.de/wiki/Non-negative_Matrix_Factorization_for_Time-Resolved_Raman_Spectroscopy_Data}}.

Typically, portal users view page content in their web browsers.
The Extension-Math processes the texvc math from the Wikitext or the Wikibase entities into HTML, allowing the browser to display the rendered formulas correctly.
This rendering pipeline leverages an external REST-based service to generate SVG images from texvc within the MediaWiki ecosystem~\cite{Schubotz2014mathoid}.

As of 2023, all Chromium-based browsers support MathML, a web standard defined by the W3C\footnote{\url{https://www.w3.org/TR/MathML/}}.
MathML is part of HTML5, akin to SVG, and enables native math rendering without needing images or browser extensions. Thus the latest versions of Edge, Chrome, Firefox, and Safari\footnote{\url{https://www.lambdatest.com/web-technologies/mathml}} fully support HTML5, including MathML since January 2023.
Due to changes in the MediaWiki ecosystem\footnote{\url{https://phabricator.wikimedia.org/T303822}} shifting away from microservices, and to augment processing capabilities for semantic annotation, the MaRDI Project developed a rendering pipeline written in pure PHP~\cite{extMathTexvc}. It generates MathML directly within the Math extension without the need for communication with REST-based services.

Initially, TexVC leverages a parsing expression grammar to extract a language-independent parse tree from LaTeX formulas. This parse tree verifies the formula's syntactical correctness and provides feedback to authors in edit mode~\cite{Stegmueller2022TexVC}. To validate this component's functionality, we used 312,190 formulas extracted from the English Wikipedia~\cite{schubotz_2014_7494266}.

The parse tree then undergoes root-first traversal, translating each item to a corresponding MathML item. The native MathML component creates the correct MathML representation for the 724 LaTeX commands supported by TexVC. The MathML used for automatic validation is generated from Mathoid~\cite{Schubotz2014mathoid} and LaTeXML~\cite{Ginev2011LaTeXML}.

The MaRDI project contributed to the open-source community by developing a PHP version\footnote{\url{https://gerrit.wikimedia.org/r/c/mediawiki/extensions/Math/+/923597}} of the TypeScript-based mhchemParser~\cite{mhchemts, Hensel2021mhchem}, integrating it into the MediaWiki ecosystem \cite{Stegmueller2023mhchem}. The \textit{mhchem} syntax is employed alongside the mathematical formula notation to write chemical equations within the Wikibase ecosystem. Implementing this component in PHP enables the processing of these equations within the PHP-based math processing pipeline. The component's functionality was validated using 116 automated tests, which incorporate the mhchem specification\footnote{\url{https://texdoc.org/serve/mhchem/0}}.

The native MathML pipeline in the MaRDI Wikibase paves the way for further advancements, such as parsing formulas from arXiv, semantic annotation of formulas, enhancing accessibility for visually impaired users, and resolving disambiguation scenarios.

\subsection{Formula Search in the MaRDI Portal}

The MaRDI Portal employs the MathSearch extension\cite{schubotz2013making}\footnote{\url{https://www.mediawiki.org/wiki/Extension:MathSearch}} to search mathematical expressions based on texvc input.
The original version utilized the Math Web Search engine, also used in the formula search of zbMATH Open \cite{muller2016full}.
In the MaRDI project, we optimized the formula search components for the Docker-based infrastructure running the latest MediaWiki versions.
The MaRDI setup\footnote{\url{https://github.com/MaRDI4NFDI/formulasearch}} uses the BaseX database to store the search index created for all formulae in the MaRDI portal.

\section{Seeding Mathematical Research Data} \label{sc.seed}


\subsection{Defining Mathematical Research Data (MathRD)}

In the realm of mathematics, research data (MathRD)~\cite{HulekMST19a, MorWiki} encompasses all data forms integral to the research process. These data forms facilitate the creation and analysis of mathematical models, substantiate proofs, test algorithms, and generally elucidate mathematical phenomena. MathRD presents itself in various types, structures, and degrees of accessibility and interoperability.

MathRD ranges from highly structured forms like symbolic or numerical data, used for encapsulating theorems, proofs, number sequences, or matrices. Geometric data, delineating objects like curves, surfaces, or polytopes, also fall into this category. In contrast, MathRD may also include less structured forms such as mathematical models or observational data. At the extreme end of the spectrum, plain text data like scientific papers, online resources, articles, and books are also encompassed under MathRD.

The MaRDI knowledge graph is engineered to effectively capture the intricate structure of this data. It does this by aggregating metadata from diverse sources, each describing mathematical research data, into a consolidated graph.

\subsection{Constructing the MaRDI Knowledge Graph}

As previously alluded to in Section~\ref{sc.portal}, the data model for the MaRDI knowledge graph is an extension of the existing data model in Wikidata. Practically, this indicates that entities for the MaRDI knowledge graph are sourced from Wikidata if pre-existing there. Once integrated into the MaRDI knowledge graph, these entities can be further enriched with additional statements, though the detailed modalities of whether and how to keep the MaRDI data in sync with Wikidata remain to be worked out.

This approach is particularly vital in the case of properties, as it necessitates the design of our data model based on the available properties in Wikidata. New properties are defined only when a corresponding one is absent in Wikidata. As a result, many of the properties found in Wikidata are also present in the MaRDI knowledge graph. We anticipate the introduction of new properties as the mathematical knowledge graph expands, which will be specifically linked to mathematics.

To facilitate entity import from Wikidata, we have established an entity importer functionality built upon the Python module WikibaseIntegrator\footnote{\url{https://github.com/LeMyst/WikibaseIntegrator}}. This module forms the basis for a Python class capable of transferring any given entity or list of entities from Wikidata to our Wikibase instance. This importer class copies not only the label, description, and aliases of the entity but also its statements, with the option to import additional connected entities. The importer class includes a parameter that determines the desired depth level of the import. By default, one level of depth is selected, importing all statements for each imported entity, and importing only label, description, and aliases for the entities mentioned in these statements.

The MaRDI knowledge graph further includes properties such as \emph{wikidata PID}\footnote{\url{https://portal.mardi4nfdi.de/wiki/Property:P11}} and \emph{wikidata QID}\footnote{\url{https://portal.mardi4nfdi.de/wiki/Property:P12}}, which store the Wikidata identifier for each imported property and item, respectively. This information is also stored in an internal SQL table, mapping the MaRDI knowledge graph identifiers to Wikidata identifiers. This table additionally incorporates a parameter for each entity indicating whether all its statements have been imported or only its label, description, and aliases.

This approach facilitates a comprehensive understanding of the overlap between the MaRDI knowledge graph and Wikidata at any given time. Moreover, it ensures synchronization between imported entities and Wikidata, and maintains consistency between the knowledge graphs, thus making the eventual integration of the MaRDI knowledge graph into Wikidata seamless.

The properties imported from Wikidata bolster the creation of new items in the MaRDI knowledge graph. In addition to the Wikidata items, metadata from eight distinct sources have been imported to varying degrees. These include formula metadata from DLMF, software metadata from swMATH, publication metadata from zbMATH Open, arXiv and crossref, metadata from resources stored at Zenodo, metadata on discrete geometric objects stored at polyDB\footnote{\url{https://polydb.org/}} and metadata on R packages published at CRAN\footnote{\url{https://cran.r-project.org/}}.

\begin{figure}[h]
\caption{MaRDI import pipeline}
\centering
\includegraphics[width=1\textwidth]{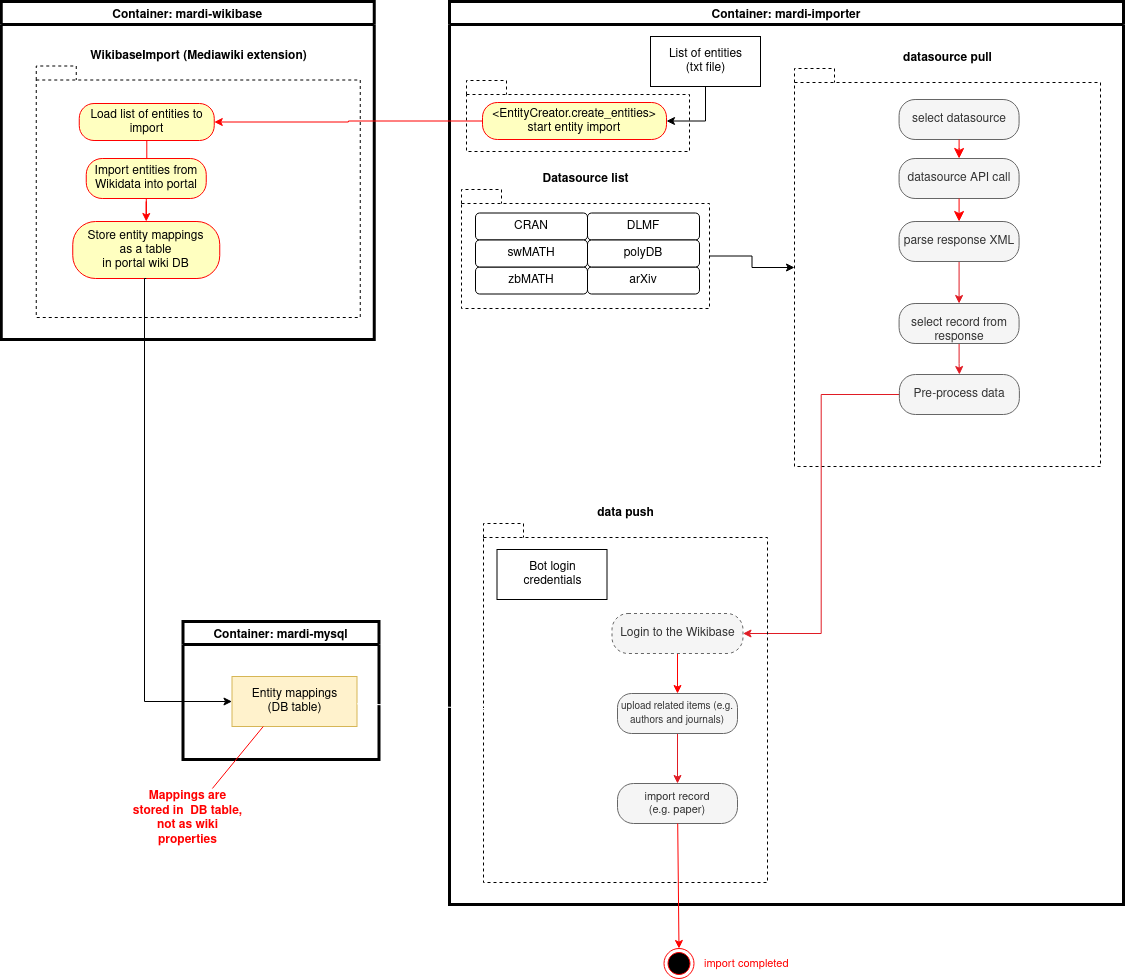}
\label{fig:workflow}
\end{figure}

The process of importing, which is displayed in \ref{fig:workflow}, occurs within the mardi-importer container, where a script is implemented for each individual datasource. Within these scripts, the data is initially retrieved. The process can be carried out either through the utilization of an Application Programming Interface (API) or by extracting information from a webpage. Subsequently, the acquired data undergoes processing, wherein relevant information is selected and it is segregated into corresponding categories such as Publications, Authors, and Journals in the case of zbMATH. The organization of the imported unprocessed data is contingent upon the manner in which it is presented inside the designated data source. In response, we have developed parsers specifically tailored to handle these variations. In certain instances, a digital object identifier (DOI) or another form of external identifier may be employed to access further information. Following this, the processed data is then uploaded to the MaRDI wikibase instance.

\subsection{Persistent Identifiers and Compliance with FAIR Principles}

Adherence to FAIR principles necessitates the assignment of persistent identifiers to each linked resource in the MaRDI knowledge graph. It's essential to differentiate between extrinsic and intrinsic identifiers:

\begin{itemize}
\item \textbf{Extrinsic identifiers}: These identifiers are linked to a specific resource through a register. DOIs, Wikibase IDs or ORCID IDs are instances of this category.
\item \textbf{Intrinsic identifiers}: These identifiers are innately linked to the resource and do not necessitate an external register for association. An example is the name of a software package, if used as its identifier.
\end{itemize}

Every resource in the MaRDI knowledge graph is assigned an extrinsic identifier by being created as a Wikibase entity. These follow the format used in Wikidata, with item identifiers beginning with 'Q' and property identifiers with 'P'. Additionally, entities imported from a given source incorporate a statement that includes the identifier from the original source. Currently, the MaRDI knowledge graph supports the following extrinsic identifiers:

\begin{itemize}
\item DOI
\item ORCID iD
\item Digital Library of Mathematical Functions ID
\item swMATH work ID
\item zbMATH Open document ID
\item zbMATH author ID
\item arXiv ID
\item arXiv author ID
\item Mathematics Subject Classification ID
\item Zenodo ID
\end{itemize}

Further, two types of resources are linked to their source using extrinsic identifiers:

\begin{itemize}
\item CRAN Project: This identifier, imported from Wikidata\footnote{\url{https://www.wikidata.org/wiki/Property:P5565}}, links to an R package published at the Comprehensive R Archive Network. The identifier is a string that corresponds to the name of the R package.
\item PolyDB ID: This identifier, created in the MaRDI knowledge graph, links to the original metadata source of a polyDB collection. The identifier string corresponds to the name of the collection.
\end{itemize}



\newcommand{\md}[1]{\href{https://portal.mardi4nfdi.de/wiki/Item:Q#1}{\texttt{Q#1}}}
\section{Case study: Orthogonal Polynomials and Special Functions} \label{sc.examples}

In this section, we showcase some capabilities of the MaRDI portal using semantically enhanced Mathematical formulas as an example, specifically focusing on Orthogonal Polynomials and Special Functions (OPSF).


In the field of OPSF, relationships between functions are essential. The NIST Digital Library of Mathematical Functions, along with its companion, the Digital Repository of Mathematical Formulae~\cite{CohlMSS14, CohlSMS15}, is an indispensable source of identifiers and formulas pertaining to this field. Their main function is as aggregators of well-established knowledge, rather than providers of novel work.

A special variant of LaTeX has been developed by the team to enable a higher level of machine-readable semantics. For example, one typically writes \verb|i| in LaTeX to represent the imaginary unit. In the context of DLMF/DRMF, however, one would use \verb|\iunit|. Both versions render the same output. However, when the LaTeX source code is web-rendered rather than PDF-generated, the \verb|\iunit| version offers the added advantage of being clickable. This improves accessibility by linking to a human-readable explanation of what the imaginary unit represents.

\begin{figure}[ht]
     \centering
           \begin{minipage}[b]{0.65\textwidth}
     \begin{subfigure}[b]{\linewidth}
         \centering
         \includegraphics[width=\textwidth]{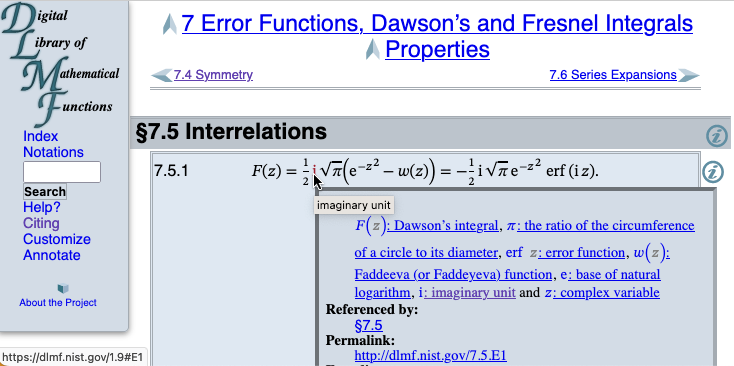}
         \caption{Screenshot from the NIST Digital Library of Mathematical Functions (DLMF) \url{https://dlmf.nist.gov/7.5} demonstrating the use of \texttt{\textbackslash iunit}.}
         \label{fig:dlmf}
     \end{subfigure}
   
     \begin{subfigure}[b]{\textwidth}
         \centering
         \includegraphics[width=\linewidth]{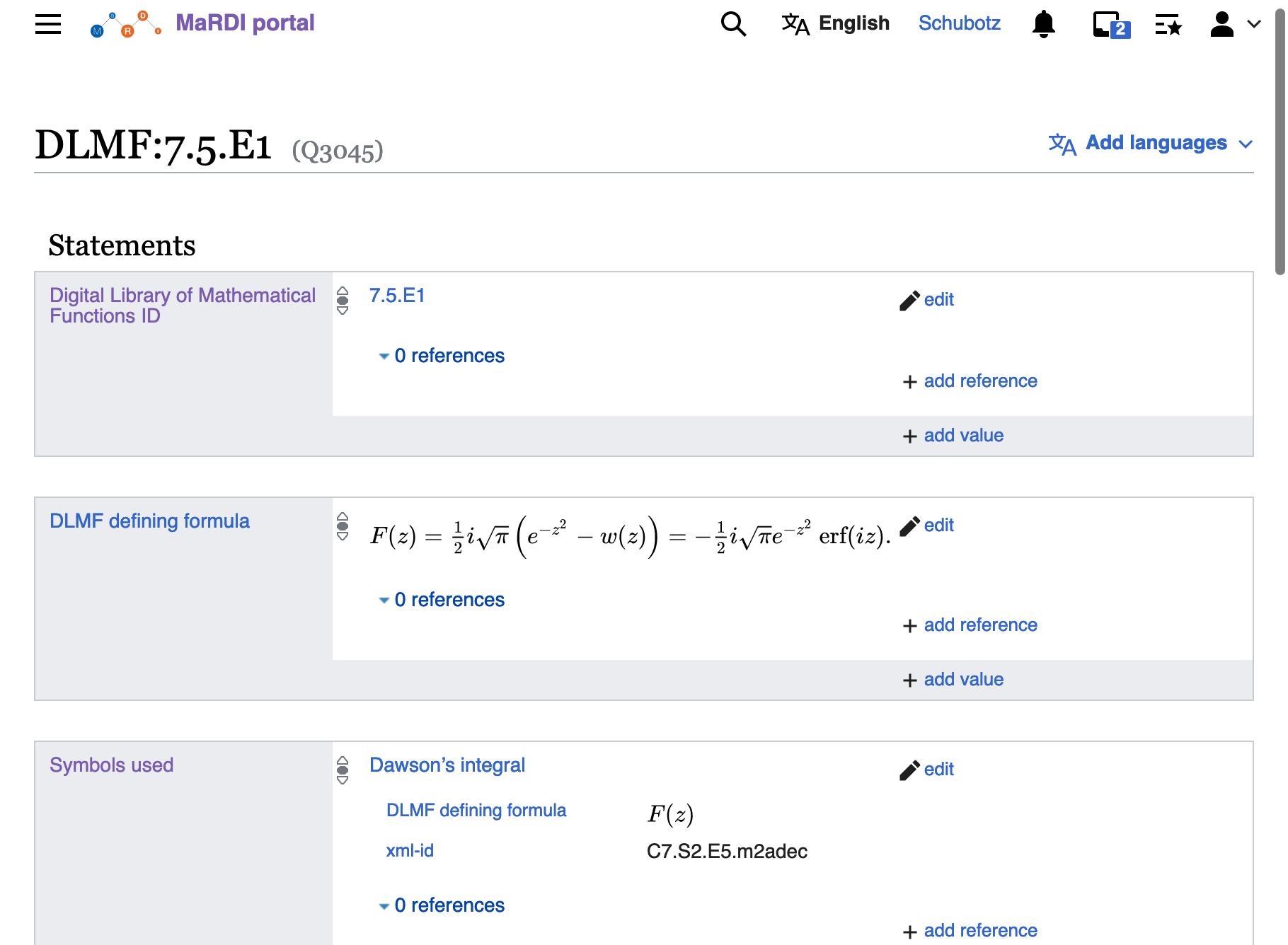}
         \caption{Representation of \ref{fig:dlmf} in the MaRDI Wikibase Instance.}
         \label{fig:three sin x}
     \end{subfigure} 
\end{minipage}      \hfill
\begin{minipage}[b]{0.32\textwidth}
    \begin{subfigure}[b]{\textwidth}
         \centering
         \includegraphics[width=\textwidth]{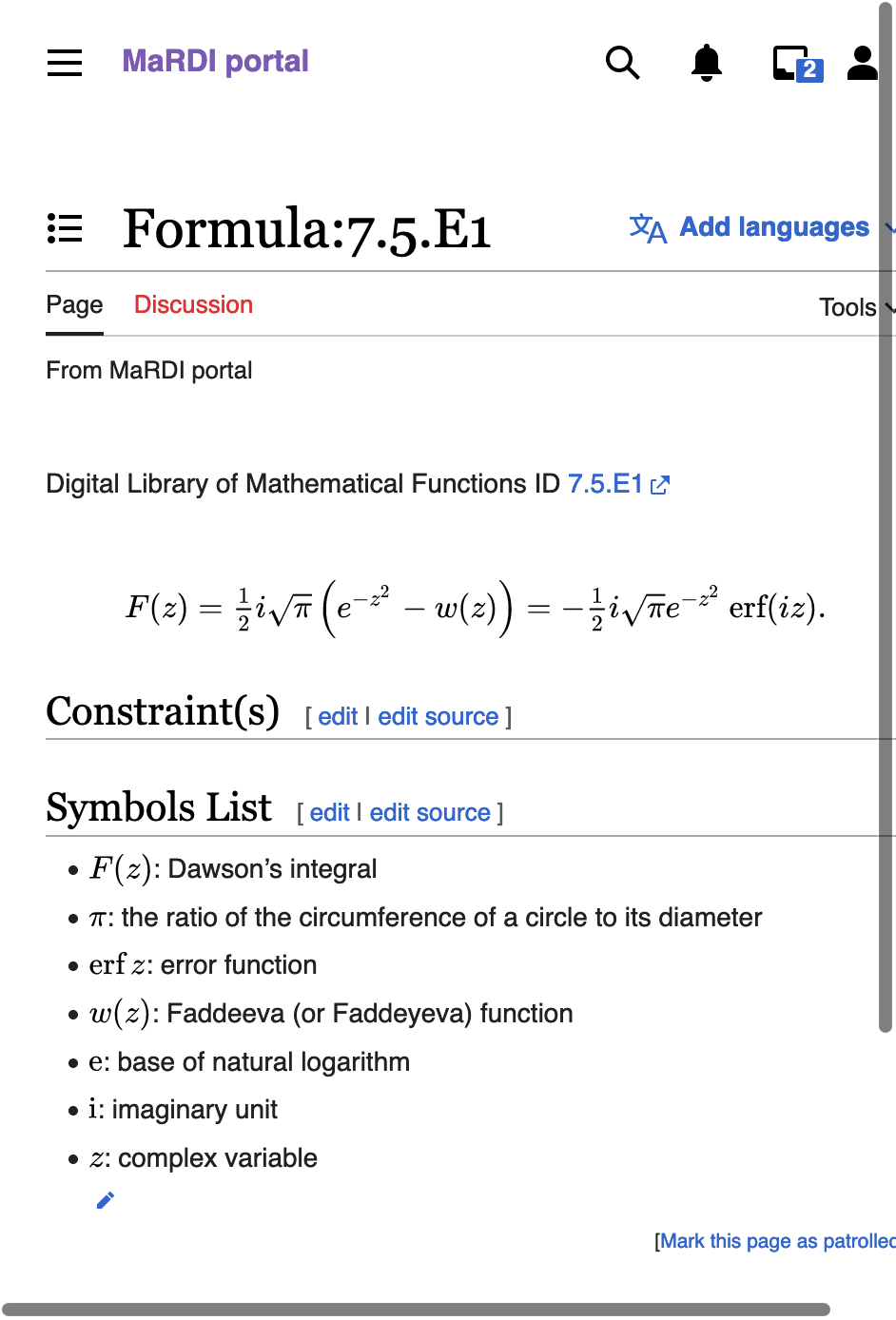}
         \caption{User friendly formula home page of \ref{fig:dlmf} in the MaRDI Portal.}
         \label{fig:y equals x}
     \end{subfigure}
     \vfill
     \vspace{2cm}     
\end{minipage}
\caption{A DLMF formula in different representations.}
\label{fig:caseStudy}
\end{figure}

In Figure~\ref{fig:dlmf}, \verb|\iunit| links to \url{https://dlmf.nist.gov/1.9#E1}, which is considered a permanent identifier. This means that the numbering remains consistent, irrespective of the introduction of additional material.

To make DLMF data accessible from the MaRDI portal, we have imported the formulas into our knowledge graph. Each formula has a corresponding Wikibase item. For example, the formula displayed in Figure~\ref{fig:dlmf} corresponds to~\md{3045} and links to the definition~\md{1399}.

Our \href{https://query.portal.mardi4nfdi.de}{SPARQL query interface} enables users to retrieve all formulas that include a link to the imaginary unit with the following command\footnote{\url{https://tinyurl.com/2bd4jv8k}}:

\texttt{SELECT ?item WHERE { ?item wdt:P4 wd:Q1399 .}}

More complex queries are also supported, such as searching for all formulas \href{https://tinyurl.com/y9c4m8sj}{indirectly dependent on the gamma function}.

We have additionally created formula homepages, similar to the DRMF project, to make the data available for classic searches and to offer a more convenient view compared to the standard Wikidata item view. Once the import of zbMATH articles is completed, bibliographic references related to the formula, software models, and algorithms will be linked from here.

Our software can also translate the expression from DLMF to Maple and Mathematica~\cite{TPAMI}. However, due to potential licensing issues, we only display the formulas on the separate site \url{https://lct.wmflabs.org}, and not in the portal.

\section{The road ahead} \label{sc.outlook}

The Bravo Release of the MaRDI platform served as a proof of concept for leveraging MediaWiki and associated services from the Wikimedia ecosystem as a platform for the National Research Data Infrastructure, focusing on the field of mathematics. We successfully demonstrated the platform's capability to handle domain-specific requirements, such as the manipulation of mathematical expressions, proving the adaptability of this general-purpose platform for specialized needs.

However, we recognize that several aspects of the platform's usability require enhancement. The current interface design, reminiscent of a Wikipedia fork, needs to be reimagined to reflect the nuances of a research data management platform. Furthermore, while advanced users might find formulating SPARQL queries straightforward, the average user might need additional support. Although Wikidata provides visual aids to improve query generation, we aim to go a step further by integrating a conversational interface to facilitate SPARQL query generation following our previous MathQA efforts~\cite{SchubotzSDN18}.

The platform's citation capabilities also leave room for improvement. While MediaWiki has long provided permanent links (using the oldid flag), these could be further enhanced. For instance, we are contemplating the implementation of a feature that stores a snapshot of the PDF on Zenodo for convenience. However, the issue of citing query results remains an open research problem, posing a challenge to reproducibility in data-intensive mathematics.

The road ahead is not without its obstacles, but we are committed to continuous improvement and development. Our aim is to deliver a platform that not only adheres to the FAIR principles but also becomes an integral part of the mathematical research data ecosystem, thereby facilitating collaboration and advancement in mathematical research.

\begin{acknowledgments}
  This work was funded by the MaRDI project~\cite{MaRDIProject} under DFG grant number 460135501. The text of this manuscript was improved with several AI tools, including Grammarly and ChatGPT.
\end{acknowledgments}

\printbibliography

\end{document}
\endinput